\documentclass[twoside,leqno,twocolumn]{article}
\usepackage{ltexpprt}
\usepackage{subfig}
\usepackage{algorithmic}
\usepackage{algorithm}
\usepackage{url}
\usepackage{times}
\usepackage{latexsym}
\usepackage{graphicx}

\newcommand{\squishlist}{
 \begin{list}{$\bullet$}
  { \setlength{\itemsep}{0pt}
     \setlength{\parsep}{3pt}
     \setlength{\topsep}{3pt}
     \setlength{\partopsep}{0pt}
     \setlength{\leftmargin}{1.5em}
     \setlength{\labelwidth}{1em}
     \setlength{\labelsep}{0.5em} } }

\newcommand{\squishend}{
  \end{list}  }

\begin{document}

\title{{Towards Breaking the Curse of Dimensionality for High-Dimensional Privacy: An Extended Version}}
\author{Hessam Zakerzadeh\thanks{University of Calgary, hzakerza@ucalgary.ca} \\
\and
Charu C. Aggrawal\thanks{IBM T.J. Watson Research Center, charu@us.ibm.com}
\and
Ken Barker \thanks{University of Calgary, kbarker@ucalgary.ca}}
\date{}

\maketitle


\begin{abstract} \small\baselineskip=9ptThe curse of dimensionality has remained a challenge for a wide
variety of algorithms in data mining, clustering, classification and
privacy. Recently, it was shown that an increasing dimensionality
makes the data resistant to effective privacy. The theoretical
results seem to suggest that the dimensionality curse is a
fundamental barrier to privacy preservation. However, in practice,
we show that some of the common properties of real data  can be
leveraged in order to greatly ameliorate the negative effects of the
curse of dimensionality. In real data sets, many dimensions contain
high levels of inter-attribute correlations. Such correlations
enable the use of a process known as {\em vertical fragmentation} in
order to decompose the data into vertical  subsets of smaller
dimensionality. An information-theoretic criterion of mutual
information is used in the vertical decomposition process. This
allows the use of an anonymization process, which is based on
combining results from multiple independent fragments. We present a
general approach which can be applied to the $k$-anonymity,
 $\ell$-diversity, and $t$-closeness models.  In the presence of inter-attribute
correlations, such an approach continues to be much more robust in
higher dimensionality, without losing accuracy. We present
experimental results illustrating the effectiveness of the approach.
This approach is resilient enough to prevent identity, attribute,
and membership disclosure attack.\end{abstract}

\vspace{-2mm}
\section{Introduction.}
The problem of privacy-preservation has been studied extensively in
recent years, because of the increasing amount of personal
information which has become available  in the context of a wide
variety of applications.  Starting with the seminal work in
\cite{sigmod2000}, a significant amount of work has been done on the
problem of privacy preservation of different kinds of data. Numerous
models \cite{sigmod2000,agrawal,chow,fli,li,machan,samarati} have been proposed for the problem
of privacy preservation. However, it has been shown that data
anonymization  is increasingly difficult with dimensionality
\cite{kanon,rand}, and the challenges extend to most privacy models.

The reason for the ineffectiveness of high-dimensional algorithms is
simple. With increasing dimensionality, a larger number of
attributes are available for background attacks,  even when the
perturbation on a single attribute is significant. As a result, it
has been shown theoretically in \cite{kanon} that significantly
larger perturbations are required with increasing dimensionality,
and this reduces the effectiveness of the approach for privacy
preservation. These results extend to a variety of models such as
$k$-anonymity and $\ell$-diversity\cite{agrawalbook}.

An important observation about a {\em blind} anonymization process
is that it  often does not (fully) recognize that dependencies among
the attributes may make a particular combination of dimensions more
or less susceptible to anonymization. While such dependencies are
{\em implicitly} utilized by many anonymization methods, their
impact is often diluted by the overall anonymization procedure.
Furthermore, the same dependencies impact the amount of
information, which may be available in a particular subset of
attributes for data mining applications. For example, an attribute
such as {\em Age} and {\em Salary} may be highly correlated, and the
{\em differential} impact of adding the attribute {\em Salary} may
be less than adding another attribute such as {\em Sex} to the data.

One solution to the curse of dimensionality is to simply use
 feature selection \cite{huai,peng} in
order to reduce the dimensionality of the data set, and retain a
small subset of attributes which retains non-redundant information
for a particular application. However, it is inevitable, that a pure
feature selection approach will lose a significant amount of
information for  many application-specific scenarios. Therefore, a
relevant question is as follows: ``{\em  Is it still somehow
possible to retain all the attributes in the data, while using the
non-redundancy of some subsets of attributes in the anonymization
process to prevent identity and attribute disclosure attack,
and also retain most of the utility in the data for mining scenarios?}''.

A less drastic approach than  feature selection is the concept of
{\em vertical fragmentation}. The idea is to break up the data set
into different subsets of attributes using vertical fragmentation,
and anonymize each subset independently. The results from the
different subsets of attributes  then need to be combined for a
particular application. Since all attributes are still retained, the
amount of information loss of fragmentation is less than that of
feature selection. The exact nature of the fragmentation may depend
upon the specific application at hand. For example:
\squishlist
\item In a supervised application, the fragments may be completely
disjoint and share no attributes other than the class attribute.
Thus, while the correspondence information among different fragments
is lost, this may not be as critical, since the class variable can
be independently learned from each fragment.  The amount of
information lost is limited in such cases,  especially if the
individual fragments are carefully chosen based on
information-theoretic principles.  The results from the different
fragments can then be combined carefully on {\em an aggregate basis}
in order to  obtain high quality classification results. Care needs
to be taken in the fragmentation process that the common class
attribute may not be used in order to partially join the fragments
together, and reduce the anonymity.
\item In an unsupervised application,  the fragments may need to
have one or more common attributes in order to ensure a limited
level of correspondence between different fragments. This case is
actually not too different from the supervised case. The main
difference is that  instead of the class attribute,  it is the
common attribute which needs to be carefully accounted for during
the fragmentation process.
\squishend
In this paper, we primarily focus on the supervised scenario of
classification as a first application.  The generalizations to other
unsupervised scenarios will be handled in future work.

It should  be emphasized that while the {\em theoretical} results of
the dimensionality curse still hold true \cite{kanon}, their {\em
practical} impact can be greatly alleviated by carefully accounting
for the nature of the  data set in a particular application.
Pathological cases may exist in which every feature  is independent
of one another, and in such cases, the earlier theoretical results
on the curse of dimensionality continue to hold true. However, such
pathological cases rarely arise in practice. Therefore, the goal of
fragmentation is to leverage the mutual information within different
features in order to  alleviate the dimensionality curse in the vast
majority of cases. The experimental results of this paper show that
the fragmentation method can achieve significant improvements over
the currently available methods. It should also be emphasized that
the fragmentation method is a {\em meta-algorithm} which can be {\em
combined with any existing anonymization algorithm} in order to
improve its effectiveness. We start with the $k$-anonymity model in
this paper because we believe such a hard problem like curse of
dimensionality must be first addressed in the simplest and most
relaxed privacy model. Then, we explain how the fragmentation
process can be generalized to satisfy the
$\ell$-diversity\cite{machan} (or $t$-closeness\cite{li}) requirement. In addition, we discuss
how the fragmentation brings in the membership disclosure
protection\cite{membershipattack}. In general, the fragmentation
process has the potential to be extended for other privacy models,
because of its meta-approach, which is more easily generalizable.
 This might eventually provide unprecedented
flexibility in using the fragmentation method  as a general-purpose
meta-algorithm in the context of a wide variety of scenarios.

This paper is organized as follows. The remainder of this section
discusses related work. Section 2 discusses the overview of the
approach for the $k$-anonymity. Section 3 discusses details of the fragment-based
$k$-anonymization method. Extending the fragmentation approach for $\ell$-diversity (or $t$-closeness) is shown in Section 4. In addition, this section shows how the fragmentation-based anonymization can prevent the membership attack. The experimental results are presented in Section 5. Section 6 contains the conclusions and summary.

\vspace{-3mm}
\section{Related Work.}
The problem of privacy preservation was first studied in
\cite{sigmod2000}. This approach was based on noise-based
perturbation of the underlying data.  Subsequently, the problem of
$k$-anonymization of data was proposed  in \cite{samarati}. Other
models for data privacy have been proposed in \cite{machan,li}.
Numerous methods
\cite{fevre,fevre2,ghinta,incognito,nonhomo,nonreciprocal} have been
proposed for effective and efficient $k$-anonymization, and this
continues to remain one of the most widely used models in the
literature because of its simplicity. The theoretical results
illustrating the degradation of privacy-preservation methods with
increasing dimensionality have been discussed in
\cite{kanon,rand,curse2}. The work in  \cite{sparsehighdim} proposed
an anonymization method for high-dimensional data. However, it is
only applicable to sparse transactional data, and is heavily
dependent of the sparse structure of transaction data in order to
achieve this goal.  Kifer \cite{injecting}  suggested  the idea of
releasing anonymized marginals beside the anonymized original table.
However, the original table must still be anonymized as a whole
which results in high information loss. Furthermore, publishing the
marginals makes the  published data highly workload dependant. The
privacy models in
\cite{setvaluedanonymization,transactionalanonymization,healthcareanonymization}
can cope with the problem of curse of dimensionality by reducing the
number of quasi-identifiers\footnote{Although the works in
\cite{setvaluedanonymization,transactionalanonymization} are
originally proposed for the set-valued (transaction) data, the
relational data can be transformed to set-valuded data for
anonymization.}. That is, they make an assumption about the number
of quasi-identifiers known by an attacker, and apply anonymization
only on limited number of quasi-identifiers. However, this
assumption may not hold true in some cases. Other relevant works
\cite{anatomy,slicing} are able to provide some  protection in the
high dimensional case, though they can be challenged in some
circumstances  \cite{kifer}. Furthermore, any of these methods can
be used in combination with our approach, which is designed as a
more general purpose meta-algorithm. Finally, the concept of
vertical partitioning and fragmentation has been explored in the
context of distributed data privacy with cryptographic protocols
\cite{vaidya}, or for capturing confidential privacy constraints in
the context of such methods
\cite{fragmentationprivacy1,fragmentationprivacy2}. The goals and
motivations of these methods are quite different, and are not at all
focused on the problem of high dimensional anonymization.

In the context of increasing dimensionality, it is natural to
explore feature selection  \cite{peng,huai} as an alternative  in
order to reduce the data dimensionality.  However,  this is a rather
drastic solution, which can impact the quality of the underlying
results significantly. Therefore, this paper proposes the approach
of fragmentation as a general-purpose meta-algorithm in order to
improve the robustness of high-dimensional  anonymization
algorithms.

\section{Overview of The Approach for The Fragmentation $K$-Anonymity}
In this section, we first introduce the most important notations used in the fragmentation-based $k$-anonymity 
in Table \ref{table:notations}. An overview is then provided about vertical fragmentation,
and its incorporation as a general-purpose
meta-algorithm for privacy preservation.  

\vspace{-8pt}
\begin{table}[H]

\caption{List of notations in fragmentation $k$-anonymity}
\setlength{\tabcolsep}{1.5pt}
\label{table:notations} \centering
\vspace{-3mm}
\begin{tabular}{|c|c|}

\hline
{\scriptsize {\bf notation}} & {\scriptsize {\bf explanation}} \tabularnewline \hline

{\scriptsize ${\cal F}$} & {\scriptsize a vertical fragmentation } \tabularnewline \hline
{\scriptsize $F_i$} & {\scriptsize $i^{th}$ fragment in fragmentation ${\cal F}$} \tabularnewline \hline
{\scriptsize $EQ_{ij}$} & {\scriptsize $j^{th}$ equivalence class in fragments $F_i$} \tabularnewline \hline
{\scriptsize $C_{ij}$} & {\scriptsize set of all class values in equivalence class $EQ_{ij}$ } \tabularnewline \hline
{\scriptsize $\mathcal P \left({S}\right)$ } & {\scriptsize power set of set $S$ } \tabularnewline \hline
{\scriptsize $|.|$} & {\scriptsize size of a set } \tabularnewline \hline

\hline

\end{tabular}

\end{table}

\vspace{-7mm}
\subsection{Vertical Fragmentation}
Let $T$ be a relation defined over a  schema $T(A_1^f,$ $A_2^f, ...,
A_n^f, A^c )$ where $A_i^f$  represents the  feature attributes, and
$A^c$ is the class attribute. A vertical fragmentation of relation
$T$ splits  the feature variables  into multiple non-overlapping
fragments. Formally, a vertical fragmentation is defined as follows:
\vspace{-0.1cm}
\begin{Definition}(VERTICAL FRAGMENTATION).
Given a relation schema $T$, a vertical fragmentation ${\cal F}$ of
$T$ is a partitioning of the attributes into  fragments ${\cal F}=\{
F_1,F_2,...,F_m \}$ such that each $F_i$ contains a  disjoint subset
of the feature attributes. Therefore, it is the case that $\forall
F_i \in {\cal F}$, $F_i \subseteq T$ and $F_i \cap F_j$ =
$\emptyset$ ($i \neq j$) and $\bigcup F_i$ = $T$ ($i$=$1,...,m$).
\end{Definition}
\vspace{-0.1cm}

In addition, it is {\em implicitly assumed} that the class attribute
$A^c$  is associated with each fragment. As we will see later, the
presence of this common attribute needs to be accounted for in a
special way since it allows the re-construction of  {\em some}
correspondence between the attribute values of different fragments.
Therefore, methods need to be designed to ensure that this
correspondence cannot be used in order to attack the anonymity of
the fragmented data. In some cases, this process requires the
perturbation of a few class values, in order to ensure
non-identifiability.

Note that the  fragmentation process is used as a meta-approach in
conjunction with an off-the-shelf anonymization algorithm.  A
fragmentation ${\cal F}$ is referred to as a \textit{$k$-anonymous
fragmentation} after applying an anonymization algorithm to it,  if
and only if the following two conditions hold:
\squishlist
\item \textbf{Fragment $k$-anonymity condition:} Each fragment in ${\cal F}$
satisfies the $k$-anonymity condition. This condition can be easily
satisfied by applying any off-the-shelf $k$-anonymity algorithm to
each fragment. \item \textbf{$K$-Anonymity non-reconstructability condition:} The
relation resulting from joining any arbitrary fragments on the class
variable satisfies the $k$-anonymity condition.
\squishend
The  number of possible fragmentations of a set of  features is
rather large in the high-dimensional case. For example, for a set of
$n$ features, there may be $O(n^n)$ possible fragmentations.
Clearly, exhaustive search through all the possible fragmentations
for the high-dimensional case may become prohibitive. Therefore, a
systematic approach is required to search the space of possible
fragmentations.  Since this work is focussed on the classification
problem, the fragmentation approach should attempt to maximize the
amount of {\em non-redundant} information contained in each feature
of a particular fragment, which is relevant for the classification
process. Correspondingly, our systematic search approach utilizes a
metric referred to as \textit{Fragmentation Minimum Redundancy
Maximum Relevance (FMRMR)} in order to create fragments.

\vspace{-3mm}
\subsection{Fragmentation Minimum Redundancy Maximum Relevance}
The ideal fragmentation is  one  in which the set of attributes in each fragment  is a comprehensive representation of the information required for the mining process.  Since this paper addresses the classification
problem, the metric will explicitly use the class variable for
quantification, though it is conceivable that the metric for other
applications would be different. In the supervised context, a
comprehensive  representation  refers to high predictability of the
class variable from the features in each fragment, while
minimizing redundancy. It is evident that  the {\em
simultaneous} incorporation of features with high mutual
information within a given fragment does not provide any additional
advantages, even when they are all highly relevant to the class
attribute. This implies that a combination of the relevance to the
class attribute and the mutual information with respect to one
another can be useful for the process of constructing a fragment.


To this effect, we draw on  the  feature selection literature, which
defines the concept of the  {\em  Minimum Redundancy
Maximum Relevance (MRMR)} metric\cite{ding,peng}. This metric uses  a dependency
quantification (denoted as $W$) among the feature variables  and a
dependency quantification  (denoted as $V$)  between the feature
variables  and the class attribute in each fragment. Our proposed
heuristic aims at maximizing the summation of MRMR for all the
fragments in a given fragmentation.  The  Fragmentation MRMR (FMRMR)
is the summation of the values of MRMR within a fragment. This value
is defined for a particular fragmentation ${\cal F}$ as follows:

\vspace{1mm}

$ FMRMR({\cal F})= \sum_{t=1}^{|{\cal F}|} (V_{t} - W_{t}) $
\vspace{1mm}

$V_{t} = \frac{1}{|A_t^f|} \sum_{j \in A_{t}^f} I(cls,j)$
\vspace{1mm}

$W_{t} = \frac{1}{|A_{t}^f|^2} \sum_{k,j \in A_{t}^f} I(k,j)$
\vspace{1mm} where:

\squishlist
\item $A_{t}^f$: set of features in fragment $t$ of fragmentation ${\cal F}$
\item $I(x,y)$: mutual information between attributes $x$ and $y$
\item $V_{t}$: total mutual information between the features
and the class attribute in fragment $t$ of fragmentation ${\cal F}$
\item $W_{t}$: total pairwise mutual information between
the features in fragment $t$ of fragmentation ${\cal F}$
\item $cls$: the class attribute
\squishend

The overall approach for the $k$-anonymity uses a  three-step technique for the
fragmentation process.  For a high-dimensional relation $T$ with $n$
features and one class attribute,  these three steps are as follows:

\begin{enumerate}
\item  Use  a carefully-designed search algorithm to
decompose the relation into  fragments. The constructed fragments
have non-overlapping sets of features together with the   class
attribute. The fragmentation process uses the afore-mentioned
measure in order to determine the optimal fragments.\vspace{-2mm}
\item Anonymize each fragment separately using an existing anonymization
algorithm, such as the Mondrian multi-dimensional $k$-anonymity
algorithm \cite{fevre}.\vspace{-2mm}
\item At this point, it should be noted that the anonymized fragments can be (partially) joined
back using the common  attribute, which in the supervised scenario
is the class attribute. Depending on the distribution of values in
the common attribute,   the result might violate the $k$-anonymity
constraint. This is essentially a  {\em $k$-anonymity non-reconstructability
condition violation}. Therefore, additional steps are required in
order to ensure non-reconstructability. The techniques for achieving
this are slightly involved and distort the class variable in such a
way that non-reconstructability is guaranteed. These methods will be
described in the next section. It should be noted that the
distortion of the class variable may result in some further
reduction in accuracy. However, in {\em practice}, for most
reasonable distributions, the required distortions are very limited,
if any.
\end{enumerate}

\vspace{-3mm}
\noindent The second step in the afore-mentioned list does not require further
explanation. Therefore,  the exposition in this paper will describe
the detailed methods for performing the first and the third steps.
For the third step, three different alternatives will be proposed.
 It should also be noted that although the class and sensitive attributes
 have been considered the same in many works,
 they might be different, and data contains many sensitive attributes in practice.
 In such cases, the other sensitive attributes also need to be fragmented
 in order to ensure that the two fragments cannot be joined.
 However, they should be fragmented only {\em after} the quasi-identifiers
  have already been fragmented (using the same approach as discussed in the next section).
   This is because it is more critical
  to ensure that quasi-identifiers are evenly distributed among fragments. 
Therefore, what follows will only focus on quasi-identifiers for simplicity.


\vspace{-3mm}

\section{Fragmentation-based $K$-Anonymization}

In this section, the first and third steps in the afore-mentioned
fragmentation meta-algorithm will be discussed. First, the fragment
construction heuristic will be introduced.
\vspace{-2mm}
\subsection{Fragment Construction Heuristic}

 As the number of features  increases, the number of possible
 fragments grows exponentially.
This explosion in the number of  fragments makes exhaustive search
in this space impractical. Therefore, we propose an algorithm which
tries to form a fragmentation with maximum $FMRMR$. For simplicity,
a binary fragmentation into two parts will be described, though it
is possible in principle to fragment  into multiple parts by
repeating the process.

We  define the \textit{FMRMR contribution} of a feature attribute
$A_i^f$ with respect to fragment $F_j$ of fragmentation ${\cal F}$ as the
difference between $FMRMR$ of ${\cal F}$ after and before adding $A_i^f$ to
$F_j$. The quantification of the mutual information between the  $n$
features and the class attribute is stored in the form of an
$(n+1) \times (n+1)$ matrix denoted by  $[MI]_{(n+1)\times(n+1)}$.

The  $FMRMR $ metric  attempts not to place features  having high
mutual information in one fragment. Therefore, as a starting point,
two features having the highest mutual information are picked as
seeds and placed in different fragments. Afterwards, in a greedy
manner, and while there exists un-assigned features, the
\textit{FMRMR contributions} of all unassigned features
with regards to both fragments are calculated. The  unassigned
feature with  the highest \textit{FMRMR contribution} is
added to the relevant partition. Finally, the common attribute (class
attribute in the supervised case)  is added to each fragment
separately.  The overall approach is illustrated in Algorithm
\ref{alg:heuristic} in the Appendix A in the supplementary materials.


\vspace{-0.3cm}
\subsection{The Final Step: $K$-Anonymity non-Recons-
tructability}
As indicated earlier,  applying a $k$-anonymity algorithm on each
fragment in order to satisfy the  \textit{fragment k-anonymity
condition} is not sufficient for ensuring non-identifiability. This
is because the common attribute (class attribute) can be used for
(very approximate) joins, and  such joins provide some additional
information about fragment correspondence. Therefore, in theory, it
may be possible that the overall anonymity level of the relation
resulted from joining $k$-anonymized fragments is less than $k$,
though in practice it is rather unlikely because of the approximate
nature of the join.



We call a fragmentation in which all fragments satisfy the
$k$-anonymity condition \textit{$k$-anonymity non-reconstructible} if the
relations resulting from joining any arbitrary fragments on the
class attribute satisfy $k$-anonymity. Similarly, a
fragmentation is called \textit{reconstructible} if  $k$-anonymity
is violated after joining some of its fragments. Definition
\ref{def:nonreconstructable} formally defines a \textit
{$k$-anonymity non-reconstructible} fragmentation.

\begin{Definition}($K$-ANONYMITY NON-RECONSTRUCTIBLE FRAGMENTATION).
\label{def:nonreconstructable} Fragmentation  ${\cal
F}=\{F_1,F_2,...,F_m\}$ which satisfies the fragment $k$-anonymity
condition is called $k$-anonymity non-reconstructible if and only if $\forall s
\in \mathcal P \left({{\cal F}}\right)$, the relation resulting from
joining members (fragments) of $s$ satisfies the $k$-anonymity
condition.
\end{Definition}

The power set $\mathcal P \left({{\cal F}}\right)$ has $2^m$
members.  However, the $k$-anonymity condition must be checked for
members of size at least 2.

It should be noted that the joining process is only approximate and
noisy, which is good for anonymization. Therefore, successful
violation attacks  of the type discussed above are often difficult
to perform in practice. For example, joining will result in some tuples that do not have any corresponding tuple in the original table. These tuples are called {\it fake tuples}, and may
sometimes be  helpful for obfuscation of identification of relevant
tuple identities.

As articulated in Definition
\ref{def:nonreconstructable}, given a fragmentation ${\cal F}$ with
$m$ fragments, the $k$-anonymity condition must be satisfied for
($2^m$-$m$-$1$) possible relations resulted from joining arbitrary
fragments. However,  the  relations resulted from joining more than
two fragments can be simply obtained by consecutive binary joins.
This paves the  way to define the $k$-anonymity non-reconstructability condition
by joining only two fragments. For instance, checking the
$k$-anonymity non-reconstructability condition on a relation resulting from
joining members of $\{F_1,F_2,...,F_n\} \in \mathcal P \left({{\cal
F}}\right)$ can be accomplished by checking the
$k$-anonymity non-reconstructability condition in each of the following binary
joins:
$I_1=F_1 \Join F_2$,   $I_2=I_1 \Join F_3$, ..., $I_{n-1}=I_{n-2}\Join F_n$.


Therefore, for simplicity,  we can continue our discussion with only
two fragments, without any loss of generality. This condition is
formally stated in Theorem \ref{thm:nonreconstructability}. An important concept to continue with the remaining of the paper is to understand the notion of equivalence classes resulting from the anonymization. In the anonymized table,
records with the same value for their quasi-identifiers constitute an \textit {equivalence class}.


\begin{theorem}($K$-ANONYMITY NON-RECONSTRUCTABILITY CONDITION). \label{thm:nonreconstructability}
The condition for a given fragmentation  ${\cal F}=\{F_1,F_2\}$
which satisfies the fragment $k$-anonymity condition, and fragments
$F_1=\{ EQ_{11},$ $ EQ_{12}$ $ ,..., EQ_{1n} \}$ and  $F_2=\{
EQ_{21}, EQ_{22},$ $ ..., EQ_{2m} \}$, to be non-reconstructible is
that one of the following must be true for each joined  pair
$EQ_{1i}$, $EQ_{2j}$:
\squishlist
\item$\sum_{c \in C_{1i} \cap C_{2j}} freq(c,EQ_{1i})\times
freq(c,EQ_{2j}) =0$
\item $\sum_{c \in C_{1i} \cap C_{2j}} freq(c,EQ_{1i})\times
freq(c,EQ_{2j}) \geq k$
\squishend
\end{theorem}

\noindent The proof of this theorem is presented in Appendix B  in the supplementary materials.


Enforcing and satisfying the
$k$-anonymity non-reconstructability condition in a fragmentation may require some
of the class values to be distorted. As explained shortly, the
change in the class values can be performed using various
strategies. However, minimizing the number of changes is always
desirable in order to retain accuracy. The design of an algorithm
which provably minimizes  the changes is computationally intractable
because of the exponential number of possibilities. Therefore, we
propose three heuristic strategies to enforce the $k$-anonymity
non-reconstructability condition. It is worth mentioning that the
utility of  each strategy is different from others. Before providing
a more detailed exposition, we introduce the concept of a {\it
dependency graph}, which provides  the logical construct necessary
for a good algorithmic design.


\vspace{-2mm}
\subsubsection{Dependency Graph}
A  {\it dependency graph} is an undirected graph structure which
captures the dependency among different equivalence classes in a
given fragmentation. Nodes in
the {\it dependency graph} are equivalence classes, and there exists
an edge between node $EQ_{ix}$ and $EQ_{jy}$ provided that:
\vspace{-2mm}
\begin{enumerate}
\item  $i \neq j $ that means $EQ_{ix}$ and $EQ_{jy}$ belong to two different fragments.\vspace{-2mm}
\item $C_{ix} \cap C_{jy} \neq \emptyset$ that means $EQ_{ix}$ and $EQ_{jy}$ have at least one class value in common.
\end{enumerate}

\vspace{-0.2cm}
 The set of all equivalence classes in a fragmentation
 may be divided into subsets having no dependency on each other. In
other words, no equivalence class from one subset can be joined with
equivalence classes in the other subset. Thus, the
\textit{dependency graph} is not connected, and the
\textit{dependency graph} components reflect the full dependencies
among all equivalence classes, rather than a single connected
\textit{dependency graph}. The process of
constructing the {\it dependency graph} is shown in
Algorithm \ref{alg:dependencygraph} in Appendix A  in the supplementary materials.


The  $k$-anonymity non-reconstructability condition is  enforced on each connected
component of the \textit{dependency graph} separately, since  there
is no inter-component dependency. The $k$-anonymity non-reconstructability
condition on each connected component can also be achieved by
enforcing it on each edge.  We introduce three different strategies
in order to achieve this goal.\\


\vspace{-0.6cm}
\subsubsection{Naive Enforcement}

Satisfying the $k$-anonymity non-reconstructability condition for an edge between
equivalence classes $EQ_{1i}$ and $EQ_{2j}$ can simply be done by
enforcing each equivalence class to have only \underline{one} class
value.  In  the naive $k$-anonymity non-reconstructability enforcement approach,
class values in each equivalence class are changed to the majority
class in that equivalence class. In this case, two given equivalence
classes either cannot be joined, or their join generates at least
$k^2$ tuples.   Such an approach is clearly suboptimal, and fails to
take full advantage of the flexibility associated with distorting
the class variable in a way which is sensitive to the behavior of
the remaining data.

\vspace{-0.3cm}
\subsubsection{Dependency Graph-based  Enforcement}

Unlike the naive approach, the   class values in {\it only}
equivalence classes violating the $k$-anonymity after being joined
are changed to the majority class value in this approach. Another
difference between this approach and the naive approach is that the
dependency graph-based approach aims at minimizing the number of
changes in each equivalence class. In order to achieve this goal,
this approach changes only one attribute in each step.

Starting from a random node (\textit{current-node}) in the {\it
dependency graph},  the \textit{dependency graph} is explored  in a
breadth-first  manner. The $k$-anonymity non-reconstructability condition is
checked between \textit{current-node} and every single unvisited
neighbor nodes. If the condition does not hold between
\textit{current-node} and one of its unvisited neighbors, the class
values with lowest frequency in the neighbor node is changed to the
majority class value until the condition is satisfied. After
satisfying $k$-anonymity non-reconstructability between $current$-$node$ and all
its neighbor nodes, $current$-$node$ is marked $visited$. This
process must be repeated for all {\it components} in the
\textit{dependency graph} until all nodes are marked $visited$. The pseudocode of this algorithm is demonstrated in Algorithm \ref{alg:dgbe} in Appendix A   in  the supplementary materials.


\vspace{-3mm}
\subsubsection{Enforcement via $\delta$-selectivity}

In spite of the approximate nature of the join between different
fragments,  they are a potential threat to  $k$-anonymity. Thus, the
prevention of violating joins is important. Publishing the class
values for {\em each and every} single tuple (row) in the anonymized
fragment is a major cause of this violation.

The $\delta$-selectivity approach changes the way in which class
values are published. This enables a more relaxed
$k$-anonymity non-reconstructability condition enforcement on the equivalence
classes. Instead of publishing the class values on a {\em per tuple}
basis, they are published on a {\em per equivalence class} basis
with the use of  {\em ambiguous values} ($slots$). In an equivalence class, each
class value has equal probability of being  assigned to a tuple.
This results in the possibility of assuming different instantiations
(or versions) for a given equivalence class. Then, given two
equivalence classes, there exist multiple ways to join them,
corresponding to different assignments of class values to tuples.
The modified $k$-anonymity non-reconstructability condition leverages this
ambiguity effectively. Appendix C.I  in the supplementary materials exemplifies tuple-level and equivalence class-level class value publishing.

As mentioned above, the  ambiguous slots in an equivalence class
$EQ_{ij}$, published at the  equivalence class-level, can take any
of the class values in $C_{ij}$. In other words, different versions
for $EQ_{ij}$ can be assumed.
\vspace{-1mm}
\begin{Definition}(EQUIVALENCE CLASS VERSION).
An arbitrary assignment of class values available in an
equivalence class $EQ_{ij}$ to ambiguous slots in $EQ_{ij}$
generates a version of $EQ_{ij}$ shown by $V(EQ_{ij})$.
\end{Definition}
\vspace{-1mm}

Although publishing the class values at the equivalence class-level
reduces the risk of $k$-anonymity violation, the resulting
equivalence classes are still vulnerable to be joined back and violate the $k$-anonymity. As an example, consider
the extreme case where the class values are unique in an equivalence
class. Each tuple in the equivalence class is assigned to a
different class value which is similar to the case in which class
values are released at the tuple level. Given two equivalence classes whose class
values are published in equivalence class-level and have at least
one class value in common, there exist different ways to join them. The number of tuples resulted from joining two equivalence classes are referred to as {\em equijoin selectivity}. Among all possible joins, those generating
$k$ tuples (or more)  are referred to as {\it
$k$-anonymity-preserving equijoins}.

\vspace{-2mm}
\begin{Definition}($K$-ANONYMITY-PRESERVING EQUIJOIN).
Given two equivalence classes $EQ_{1i}$ and $EQ_{2j}$ whose class
values are published at the equivalence class level, and which share
at least one class value,
 the join between
V($EQ_{1i}$) and V($EQ_{2j}$) is a $k$-anonymity-preserving
equijoin if and only if  it produces at least $k$ tuples.
\end{Definition}

\vspace{-2mm}

We can now define the equijoin selectivity privacy level in terms
of the possible equijoins between two equivalence classes.

\vspace{-2mm}
\begin{Definition}(EQUIJOIN SELECTIVITY PRIVACY LEVEL).
The ratio of number of $k$-anonymity-preserving equijoins in joining
two equivalence classes $EQ_{1i}$ and $EQ_{2j}$ to the total number
of possible equijoins in joining the same equivalence classes is
referred to as equijoin selectivity privacy level  of $EQ_{1i}$ and
$EQ_{2j}$. This value is denoted by $ \eta(EQ_{1i},EQ_{2j})$. In
other words, $ \eta(EQ_{1i},EQ_{2j}) =$
\vspace{2mm}

$ \frac{|ds_p=\{\{ V(EQ_{1i})\} \Join
\{V(EQ_{2j})\} |\hspace{1mm}|\{ V(EQ_{1i})\} \Join
\{V(EQ_{2j})\}| \geq k \}|} {|ds_w=\{\{V(EQ_{1i})\} \Join
\{V(EQ_{2j}) \} |}.$
\end{Definition}

Intuitively, $\eta(EQ_{1i},EQ_{2j})$ indicates the probability that
the result of joining $EQ_{1i}$ and $EQ_{2j}$  is a $k$-anonymous
equivalence class. As an example, when the value of  $\eta$ is  1,
it  indicates that $\forall v_1 \in {V(EQ_{1i})}, v_2 \in
{V(EQ_{2j})}$, we have  $|v_1 \Join v_2|\geq k$.  In other words,
all possible instantiations will result in a
$k$-anonymity-preserving equijoin.

\vspace{-2mm}
\begin{Definition}($\delta$-SELECTIVE $K$-ANONYMOUS FRAG.).
A fragmentation ${\cal F}=\{F_1,F_2\}$ that satisfies the
\textit{fragment $k$-anonymity condition} is called $\delta$-selective
if and only if  $\forall EQ_{1i}, EQ_{2j}$ we have
$\eta(EQ_{1i},EQ_{2j}) \geq \delta$.
\end{Definition}
\vspace{-2mm}

Algorithm \ref{alg:deltaselectivity} in Appendix A   in the supplementary materials shows how $\delta$-selectivity
can be enforced on a fragmentation ${\cal F}$.


\vspace{-0.3cm}

\section{Extension to $\ell$-Diversity}
Analogous to the $k$-anonymity case, a fragmentation is called an
$\ell$-diverse fragmentation, if and only if 1) each fragment
satisfies the $\ell$-diversity requirement (fragment
$\ell$-diversity condition) and 2) joining the fragments does not
violate the $\ell$-diversity requirement ($\ell$-diversity
non-reconstructability condition). Satisfying the fragment
$\ell$-diversity condition is similar to that of the $k$-anonymity
case. However, for  $\ell$-diversity, the  non-reconstructability
condition is different. Table \ref{table:notationsdiversity}
describes the notations used in this section. A brief overview of
the steps for the fragmentation-based $\ell$-diversity is provided
below:
\vspace{-4mm}
\begin{table}[H]

\caption{List of notations in fragmentation $\ell$-diversity}
\vspace{-2mm}
\setlength{\tabcolsep}{1.5pt} \label{table:notationsdiversity}
\centering

\begin{tabular}{|c|c|}

\hline
{\scriptsize {\bf notation}} & {\scriptsize {\bf explanation}} \tabularnewline \hline

{\scriptsize ${\cal F}$} & {\scriptsize a vertical fragmentation } \tabularnewline \hline
{\scriptsize $F_i$} & {\scriptsize $i^{th}$ fragment in fragmentation ${\cal F}$} \tabularnewline \hline
{\scriptsize $S_i$} & {\scriptsize $i^{th}$ segment} \tabularnewline \hline
{\scriptsize $CK_{ij}$} & {\scriptsize data chunk belongs to segment $S_i$ and fragment $F_j$} \tabularnewline \hline
{\scriptsize $EQ_{ij}^s$} & {\scriptsize $i$th equivalence class belongs to chunk $CK_{sj}$} \tabularnewline \hline
{\scriptsize $C_{ij}$} & {\scriptsize set of class values for $CK_{ij}$} \tabularnewline \hline
{\scriptsize $C_{ij}^s$} & {\scriptsize set of class values for $EQ_{ij}^s$} \tabularnewline \hline
{\scriptsize $l_i$} & {\scriptsize diversity level of segment $S_i$} \tabularnewline \hline
{\scriptsize $|.|$} & {\scriptsize size of a set } \tabularnewline \hline

\hline

\end{tabular}

\end{table}
\vspace{-7mm}

\begin{enumerate}
\item Use the fragment construction algorithm
proposed in Section 4.1 solely to {\em compute} the
best fragmentation ${\cal F}=\{F_1,F_2,...,F_n\}$. However, the
fragmentation is not actually {\em executed} in this step.
\vspace{-3mm}
\item Cluster data records
into $m$ segments ${\cal S}=\{S_1,S_2,...,S_m\}$ using a top-down
clustering algorithm. Stop dividing each segment $S_i$ into further
sub-segments if the resulting sub-segments either violate the
$k$-anonymity or $\ell$-diversity requirement.  Final segment $S_i$
has the diversity level $l_i$.
\vspace{-3mm}
\item Vertically partition each
segment $S_i$, using the fragments found in Step 1, into $n$ data chunks.
The diversity level of each data chunk $CK_{ij}$ is equal to the diversity level of segment $S_i$, which is $l_i$.
\vspace{-3mm}
\item Use any off-the-shell $\ell$-diversity algorithm to anonymize
each chunk $CK_{ij}$. However, the diversity requirement of  each
$CK_{ij}$ must be set to $l_i$.
\vspace{-6mm}
\item Merge equivalence classes belonging to the same vertical fragments and publish them as one fragment.
\vspace{-1mm}
\end{enumerate}

In the afore-mentioned algorithm, it is worth noting that  the
clustering algorithm in the second step should be adjusted to the
relevant workload. Since the  workload  in our approach is
classification, a classification-oriented clustering algorithm
results in higher utility.

The $\ell$-diversity non-reconstructability condition may be
violated if at least two chunks $CK_{si}$ and $CK_{sj}$ are joined
and the resulting data set violates either the $k$-anonymity or the
$\ell$-diversity requirement. We prove that this will not be the
case according to the way equivalence classes in chunks have been
formed. For simplicity, we prove this for the case of  two vertical
fragments. However, the result is true in general.


\vspace{-1mm}
\begin{theorem} The data set resulting from
joining chunks $CK_{s1}$ and $CK_{s2}$ neither violates
$k$-anonymity nor $\ell$-diversity.


\end{theorem}

\vspace{-3mm}
 Again, the proof of this theorem is presented in Appendix B   in the supplementary materials for the sake of brevity. It is easy to show that this extension can be utilized for $t$-closeness by simply enforcing $t$-closeness instead of $\ell$-diversity in the aforementioned steps.

\vspace{-3mm}
\subsection{Membership Disclosure Protection}
Fragmenting the data can help protect against
 membership attack\cite{membershipattack} by disassociating different attributes.
 As discussed in \cite{membershipattack},  the ability to determine
  presence or absence of a  subject's record
 in the published data is a privacy threat.
 This  can be  done by comparing the subject's
  quasi-identifiers with the published quasi-identifiers.

Consider an attacker trying to find out the membership of subject
$v$ in the published fragmented data. As attributes are fragmented,
the attacker must find the matching equivalence class in each
fragment to which the subject's attributes belong. This may not be
possible considering the generalization applied on the attributes.
However, provided that the attacker succeeds in finding the matching
equivalence classes $EQ_{1i}$, $EQ_{2j}$,..., $EQ_{np}$, the
likelihood that the record pertaining to $v$ exists in the published
fragmented data is $\frac{|EQ_{1i}\Join EQ_{2j} \Join ...\Join
EQ_{np}|}{|F_1 \Join F_2 \Join ... \Join F_n|} = \frac{|EQ_{1i}\Join EQ_{2j} \Join ...\Join
EQ_{np}|}{\sum_{i} \sum_{j}...\sum_{p} |EQ_{1i}\Join EQ_{2j} \Join ...\Join
EQ_{np}|} $.

\noindent This likelihood is mostly impacted by $p$ (the number of
vertical fragments), anonymity level (either $k$ or $\ell$), and
$|D|$ (size of the data set).  In most cases, the value of  $p$ is
small, and $|D| \gg k$. Therefore,  the numerator of the likelihood
formula becomes much smaller than the denominator.  When the value
of  $p$ increases, the number of common class values among specific
equivalence classes $EQ_{1i}$, $EQ_{2i}$,..., and $EQ_{np}$ drops
and even tends to zero in many cases. Therefore, the chance of a
successful membership attack becomes negligible. In general, the
fragmentation-based anonymization provides strong protection against
membership attack.



\vspace{-4mm}

\section{Experimental Results}
In this section, we will present the experimental results showing
the effectiveness of our method. The goal is to show that the
fragmentation process is able to retain greater utility of the data
both in terms of classification measures and information loss
measures, at the same level of privacy.

We utilized two metrics to evaluate the effectiveness of our
proposed method, {\em information loss} and {\em weighted F-measure}
to capture the total amount of lost information  and evaluate the
utility of the data anonymized by our meta-algorithm, respectively.
Details on how these metrics are used on fragmented data are
available in Appendix D.I   in the supplementary materials. \vspace{-2mm}
\subsection{Baselines}
Since the goal was to show the effectiveness of the fragmentation
approach as a meta-algorithm,  the baseline for the approach were
the results for the anonymization process with and without
fragmentation. The Mondrian multidimensional anonymity method
\cite{fevre} was used for {\em both} the fragmented, {\em and}
the unfragmented scenario. More accurately, in Step 2 of fragmented $k$-anonymity, we used $median$ Mondrian and in Step 4 of fragmented $\ell$-diversity, we utilized $\ell$-diversity Mondrian.  Therefore, the qualitative improvements
show the  effects of fragmentation,  as a methodology to improve the
effectiveness of an off-the-shelf approach.
\vspace{-2mm}
\subsection{Data Sets}
Real  data set \textit{Musk}  from the {\em  UCI Machine Learning
Repository}\footnote{\url{http://archive.ics.uci.edu/ml.}} was
used. The detailed description of the data set is provided in the Appendix D.II of the supplementary materials.

\vspace{-2mm}
\subsection{Results}
\label{resultspart}
In each case, the results were measured with varying dimensionality
and anonymity level.  In each case, the anonymity level was varied
after fixing the dimensionality, and the dimensionality was varied
on fixing the anonymity level.  The anonymity level was fixed to 40, when the dimensionality was varied on the
$X$-axis.  While varying the anonymity level on the $X$-axis, the
dimensionality was fixed to 40. In case of $\ell$-diversity, the value
of  $l$ is set to 2. In addition, we fixed $\delta$ to 0.5 in
$\delta$-selective enforcement approach. It is important to note
that the information loss results do {\em not} vary with the
different strategies for ensuring $k$-anonymity
non-reconstructability, which affect only the class variable. Since
the information loss metrics are based on the feature variables
only, a single chart will be shown for the case of information loss,
whereas the performance results for different approaches of
$k$-anonymity (based on different strategies for ensuring
$k$-anonymity non-reconstructability) and $\ell$-diversity will be
shown in the case of F-measure separately by means of solid and
dashed lines, respectively.

\begin{figure}
\vspace{-1.1cm}
\centering \caption{Information loss vs. dimensionality \& $k$}\vspace{-4mm}
\subfloat[\textit{Info loss vs. dim}]{
\includegraphics[scale=0.2]{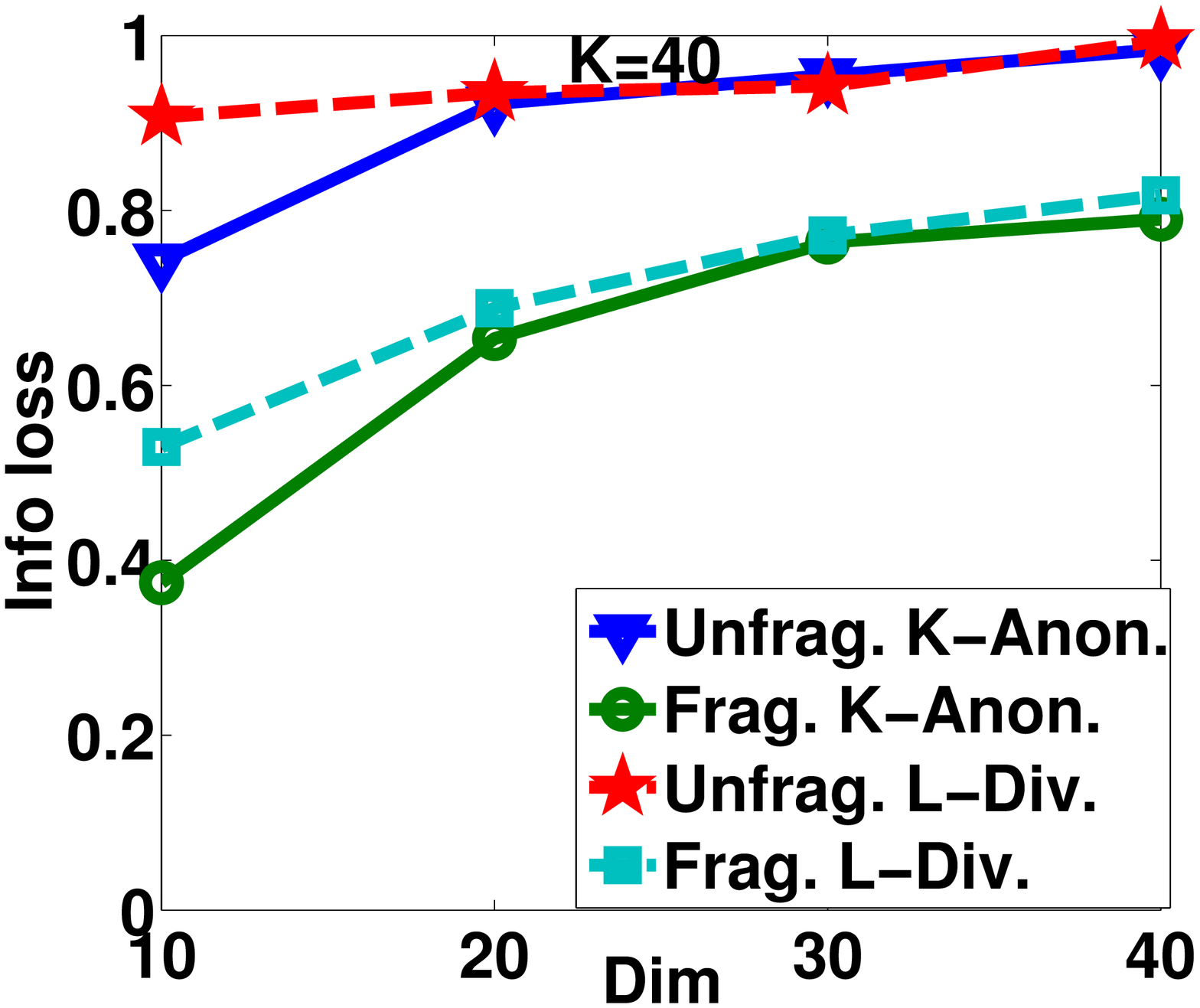}
}\subfloat[\textit{Info loss vs. $k$}]{
\includegraphics[scale=0.2]{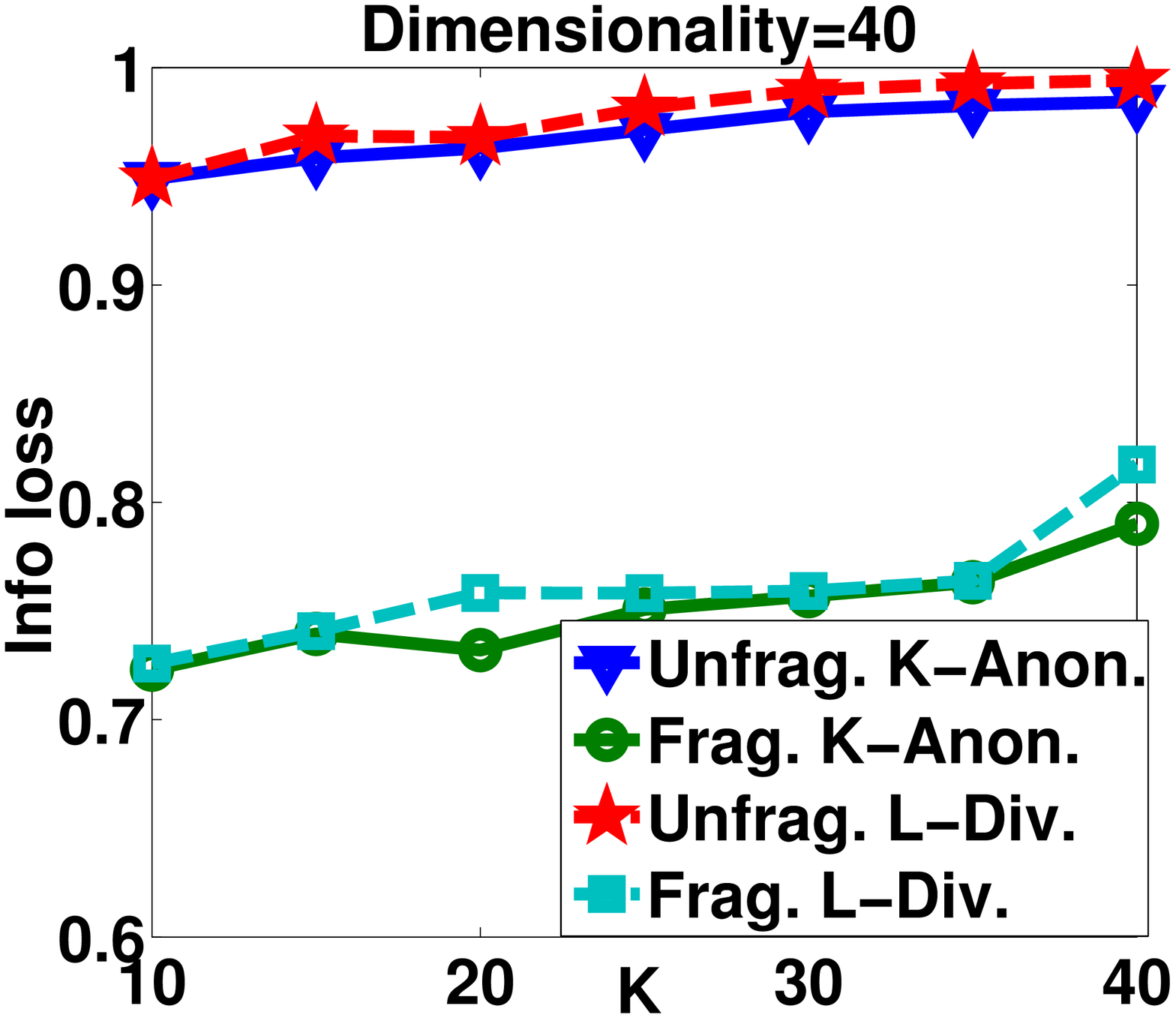}
}
\label{fig:infolossdim}
\vspace{-1.1cm}
\end{figure}

The information loss with varying dimensionality is illustrated in Figure
\ref{fig:infolossdim}a.  The dimensionality is illustrated on the
$X$-axis, and the information loss is illustrated on the $Y$-axis in
each case. Besides, diversity level is set to 2 for $\ell$-diversity in all
experiments. It  is evident that the information loss of the {\em
unfragmented} approach (for both $k$-anonymity and
 $\ell$-diversity) increases with increasing dimensionality,
which is in agreement of the results found earlier in \cite{kanon}. In
fact, the error touches almost its upper bound, which implies that
each generalized value  starts losing more and more of its
specificity in the unfragmented case.  On the other hand, the
fragmentation method shows drastic improvements in the amount of
information loss, for both $k$-anonymity and $\ell$-diversity.
This implies that a significant amount of
attribute specificity is retained in each  fragment.

 The  information loss  with increasing anonymity level
 is illustrated in Figure \ref{fig:infolossdim}b.
 The anonymity level is illustrated on the $X$-axis, whereas the
 information level is illustrated on the $Y$-axis.
It is  not surprising that the information loss increases with the
anonymity level, and enforcing diversity. However, as in the case of the results with
increasing dimensionality, the improvements achieved by
fragmentation were significant.

The comparisons for the F-measure with increasing
dimensionality are much more tricky. This is because the addition of
more dimensions to a data set affects the classification precision and recall (and hence F-measure) of
the data in two mutually contradictory ways:
\squishlist
\item A larger number of dimensions provides greater knowledge (in terms of more attributes) to
the classifier in order to improve its precision and recall. \item Data sets of
larger dimensionality will have greater information loss on a {\em
per attribute basis}, and this reduces the effectiveness of the
classifiers.
\squishend

\begin{figure}
\vspace{-0.9cm}
\centering \caption{Prediction F-measure on $Musk$ vs.
dimensionality} \vspace{-4mm}\subfloat[\textit{J48 classifier}]{
\includegraphics[scale=0.2]{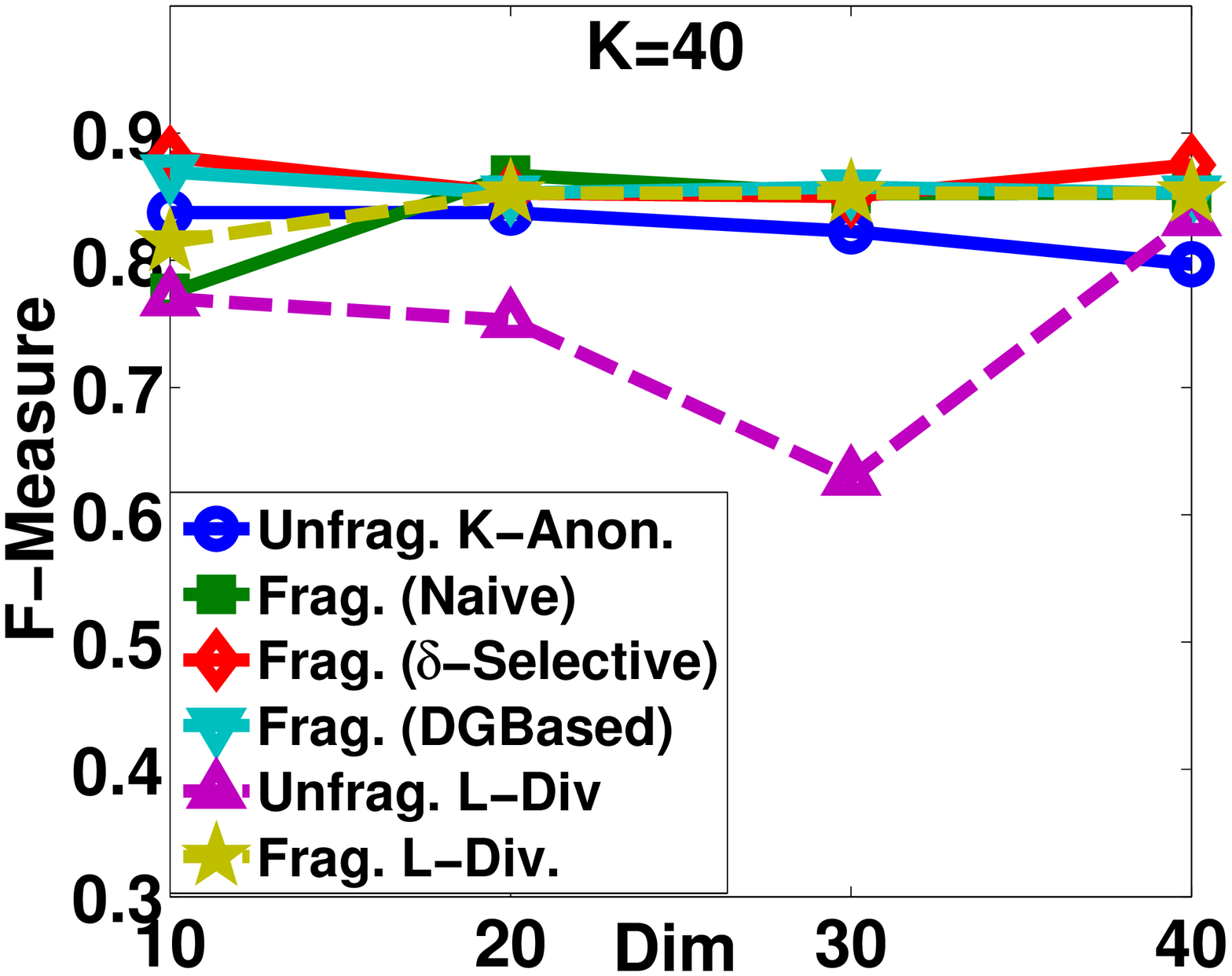}
}\subfloat[\textit{k-NN classifier}]{
\includegraphics[scale=0.2]{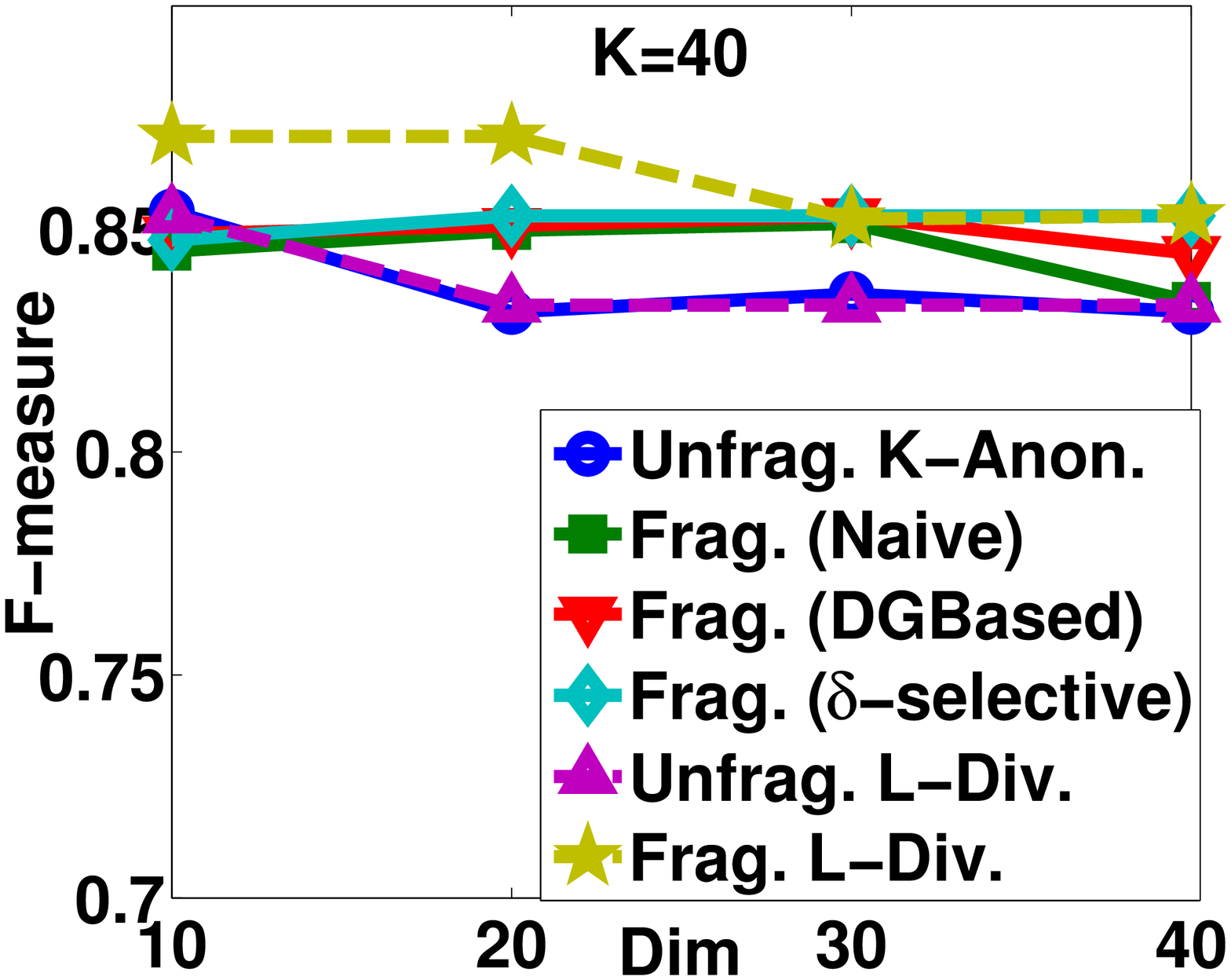}
}
\label{fig:muskaccuracydim}
\vspace{-0.9cm}
\end{figure}

So how does this tug-of-war between two mutually contradictory
effects impact the final classification results, and how does the
fragmentation process affect this tradeoff?
 Figure
\ref{fig:muskaccuracydim} compares the
prediction F-measure of  the unfragmented and fragmented
anonymization methods  with increasing dimensionality for two
classifiers.
In all cases, the different variants of the fragmentation scheme
have higher classification accuracy than the unfragmented scheme.
  Even the naive scheme (in fragmented $k$-anonymity) was often able
to perform better than the unfragmented approach in spite of its
relative lack of sophistication in performing the class distortions.
The difference in F-measure becomes even more drastic in case of
$\ell$-diversity and the fragmented anonymization achieves up to
28\% improvement over unfragmented scenario.

It is  also immediately evident that the trend
with increasing dimensionality is specific to the choice of data
set, classifier, and specific approach.  In particular, an
interesting trend is that the F-measure changes only a little bit
with increasing dimensionality in many cases, especially for
unfragmented data. This is because the anonymization process in the
unfragmented case changes values for many feature attributes to the
very general value  as the dimensionality grows. This change turns a
high-dimensional data set into a data set with very few useful
features for the classification. This phenomenon is reflected in the
F-measure of unfragmented anonymization shown in Figure
\ref{fig:muskaccuracydim}, which often
does not vary much. In fact, only 6 feature attributes in $Musk$ played a significant role in  the
classification. As a result, the F-measure does not vary too much
with increasing dimensionality. Besides, as the equivalence classes
have different class labels in the case of $\ell$-diversity, the
precision and recall degrade dramatically and cause the F-measure to
be very low for unfragmented $\ell$-diversity.

\begin{figure}
\centering
\caption{Prediction F-measure on $MUSK$ vs. $k$}
\vspace{-4mm}
\subfloat[\textit{J48 classifier}]{
\includegraphics[scale=0.2]{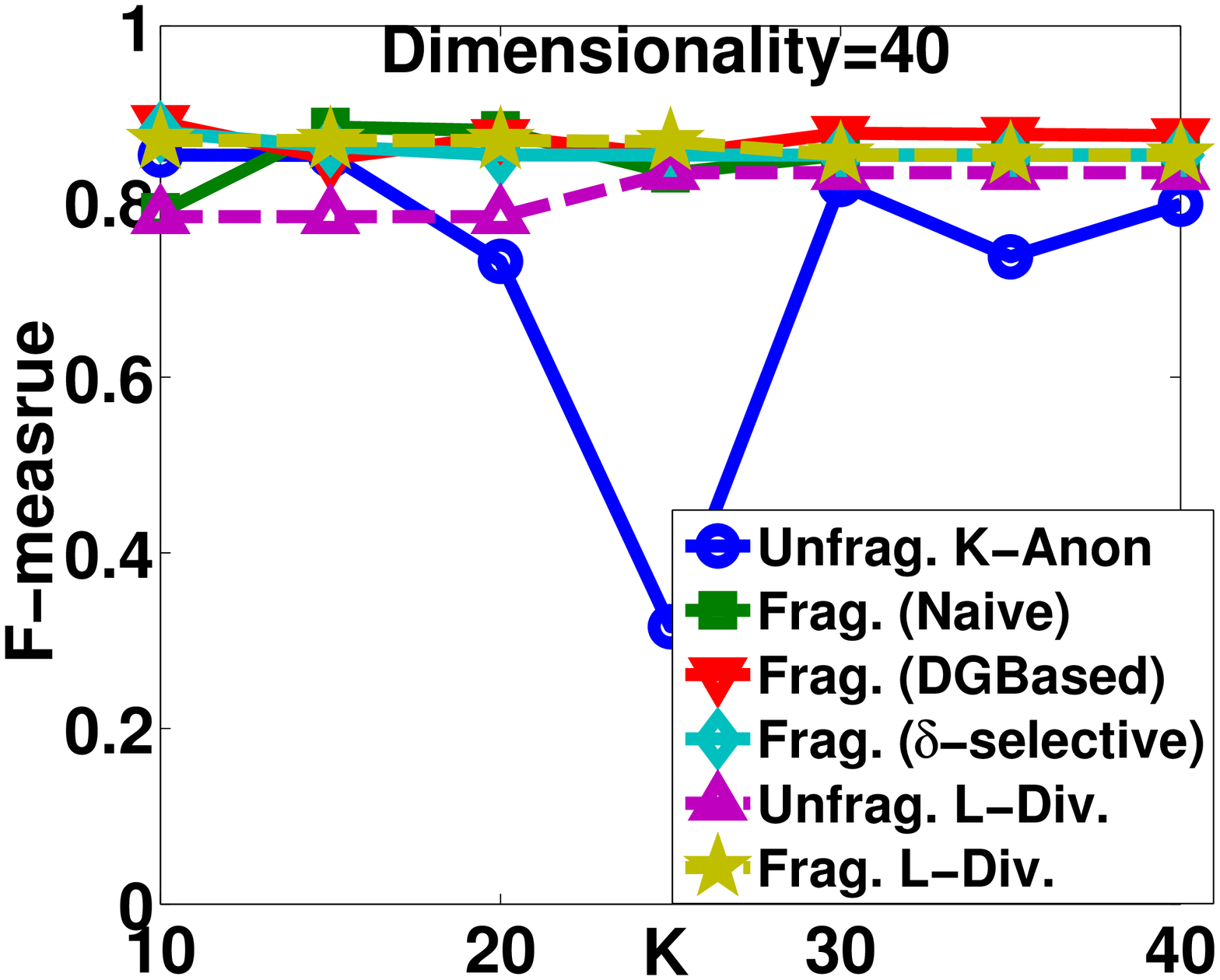}
}\subfloat[\textit{k-NN classifier}]{
\includegraphics[scale=0.2]{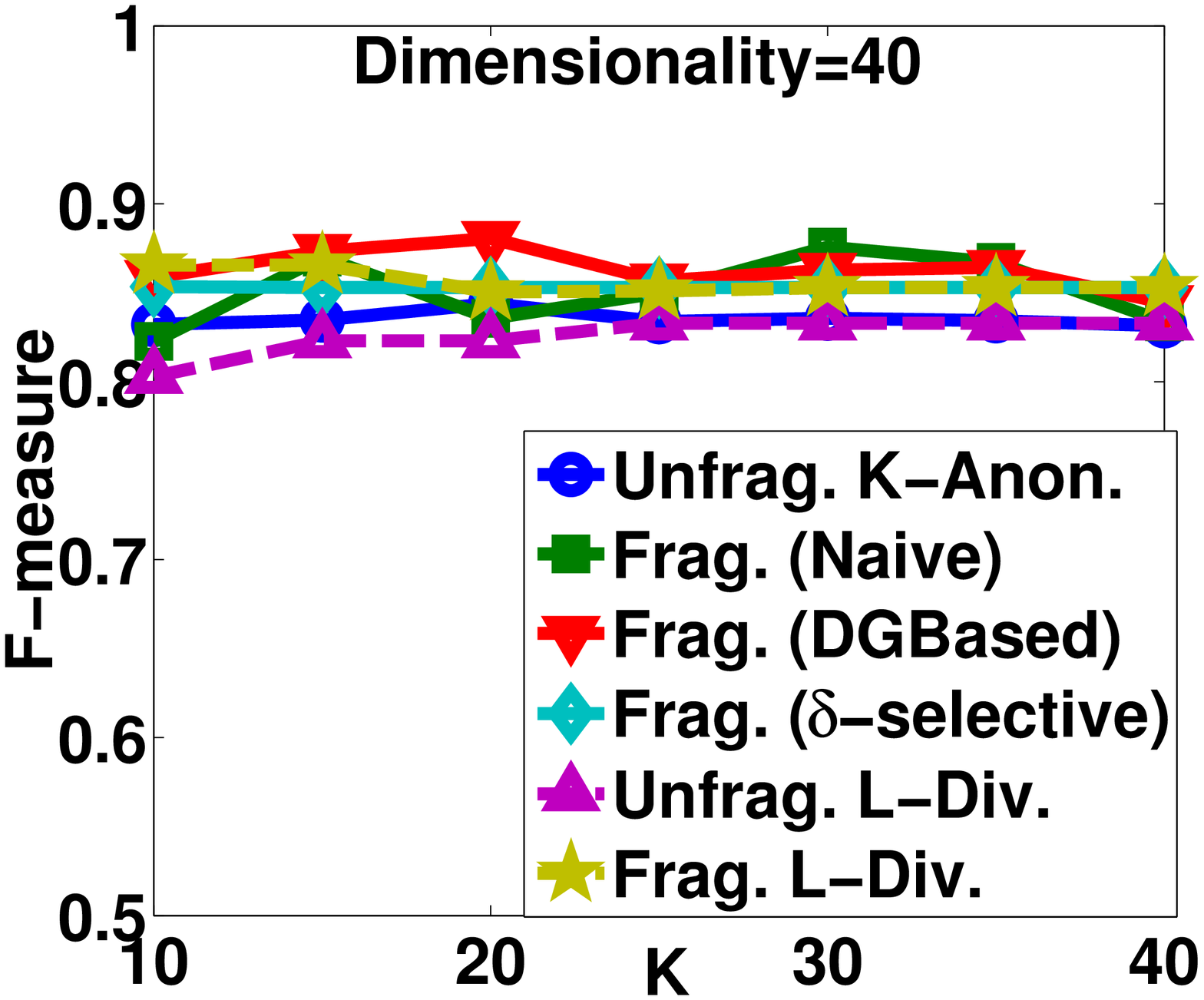}
}
\label{fig:muskaccuracyk}
\vspace{-1cm}
\end{figure}

The effect of $k$ on classification F-measure is illustrated in   Figure \ref{fig:muskaccuracyk}. The fragmented
anonymization reveals a prominent improvement of  up to 54\%
compared to the unfragmented anonymization. Normally, we expect the
prediction F-measure to decline with increasing values of $k$. While
this was  often the case, there were also a few cases, where  it has
an unexpected rise. This trend has also sometimes  been observed in
earlier work, and is  a result of the aggregation effects of the
anonymization procedures (sometimes) removing the noisy
artifacts in the data.

\vspace{-2mm}

\vspace{-1mm}
\section{Conclusions and Summary}
This paper presents a method for fragmentation-based anonymization
for high-dimensional data. While the curse of dimensionality  is a
fundamental theoretical barrier, it is often possible to  obtain
effective results in practice. This paper uses fragmentation as a
general purpose methodology to improve the effectiveness of any
off-the-shelf algorithm for the anonymization process. Experimental
results show significant improvements of the utility of the data
after the fragmentation  process. This meta-algorithm approach is
fairly general and has the potential to be extended to a wider
variety of scenarios and privacy models and workloads. This will be the focus of our future work.

\clearpage

\section*{Supplementary Materials}
\section*{Appendix A. Pseudocodes}

Algorithm \ref{alg:heuristic} depicts how a high-dimensional data set can be broken down into fragments.
\begin{algorithm}
\caption{Fragment Construction Algorithm}          
\label{alg:heuristic}
\begin{algorithmic}[1]                    

\STATE // \textit{n} denotes the number of features

\STATE {\textit MI}: a ($n+1$) by ($n+1$) matrix storing the mutual information (between the features and the class attribute)

\STATE $Fragment1$, $Fragment2$, $nonAssignedFs$= {\bf empty}

\STATE Select two features having the maximum mutual information as two seeds, $seed1$ and $seed2$.
\STATE $Fragment1$.add($seed1$)
\STATE $Fragment2$.add($seed2$)
\STATE Add the rest of features to $nonAssignedFs$

\STATE {\bf while} ($nonAssignedFs$ not {\bf empty})
\STATE  \hspace{3mm} $bestF$,$bestFragment$={\bf empty}
\STATE  \hspace{3mm} $maxContribution$=0
\STATE  \hspace{3mm} {\bf foreach} $a_i$ in $nonAssignedFs$
\STATE  \hspace{5mm} {\bf foreach} $f$ in {$Fragment1$,$Fragment2$}
\STATE  \hspace{7mm} {\bf if}($contribution$($a_i$,$f$)$>maxContribution$)
\STATE  \hspace{9mm} $bestF$=$a_i$
\STATE  \hspace{9mm} $bestFragment$=$f$
\STATE  \hspace{9mm} $maxContribution$=$contribution$($a_i$,$f$)
\STATE  \hspace{3mm} Add $bestF$ to $f$
\STATE  \hspace{3mm} Remove $bestF$ from $nonAssignedFs$
\STATE  Add the class attribute to $Fragment1$ and $Fragment2$
\end{algorithmic}
\end{algorithm}

\begin{algorithm}[H]
\caption{CreateDependencyGraph($set$-$of$-$all$-$EQ$)}          
\label{alg:dependencygraph}                           
\begin{algorithmic}[1]                    

\STATE  mark all equivalence classes in $set$-$of$-$all$-$EQ$ as $un$-$visited$
\STATE  $c$-$id$=1
\STATE  $toBeProcessed$={\bf empty}
\STATE  $root$=random($set$-$of$-$all$-$EQ$) // choose an unvisited equivalence class randomly
\STATE  $toBeProcessed$.enqueue($root$)
\STATE  {\bf while} ($toBeProcessed$ not {\bf empty})
\STATE  \hspace{3mm} $current$-$node$=$toBeProcessed$.dequeue()
\STATE  \hspace{3mm}  visit($current$-$node$)
\STATE  \hspace{3mm} {\bf foreach} $EQ$ in dependant($current$-$node$)
\STATE  \hspace{6mm} {\bf if}($EQ$ not visited)
\STATE  \hspace{9mm} draw an edge from $current$-$node$ to $EQ$ in
\STATE  \hspace{9mm} component with id=$c$-$id$
\STATE  \hspace{9mm}  $toBeProcessed$.enqueue($EQ$)
\STATE  {\bf if} (any un-visited equivalence class left)
\STATE  \hspace{3mm} $c$-$id$++
\STATE  \hspace{3mm} go to line 4
\end{algorithmic}
\end{algorithm}

Algorithm \ref{alg:dependencygraph} takes the set of all equivalence
classes in a given fragmentation, and constructs the \textit{dependency graph}.
At the end of this process, the variable
$c$-$id$ indicates the number of components with no dependency
(connection). The subroutine $dependant$ takes an equivalence class
$eq$  in either $F_1$ or $F_2$,  and returns the set of equivalence
classes in the other fragment that can be joined by $eq$.

Algorithm \ref{alg:dgbe} shows the procedure for enforcing the
condition in Theorem \ref{thm:nonreconstructability} in a {\it
dependency graph}.  This algorithm is invoked for each  component the
\textit{dependency graph} in order to ensure non-identifiability.
In Algorithm \ref{alg:dgbe}, the subroutine $change$ (line 9)
changes a class value having the lowest frequency in its equivalence
class to the majority class value in that equivalence class.
However, this change might affect the previously satisfied
$k$-anonymity non-reconstructability condition between node $un$ and its $visited$
neighbors. Thus, lines 10-14 re-check the previously satisfied
reconstructability condition, and if needed, the  subroutine
$change$ is called  as necessary in order  to re-satisfy it.

\begin{algorithm}
\caption{DGBE($d$-$graph$)}          
\label{alg:dgbe}                           
\begin{algorithmic}[1]                    
\STATE $processing$-$queue$ = {\bf empty}
\STATE mark all nodes in $d$-$graph$ as $unvisited$
\STATE $processing$-$queue$.enqueue(random($d$-$graph$))
\STATE {\bf while} ($processing$-$queue$ {\bf not empty})
\STATE \hspace{3mm} $current$-$node$ = $processing$-$queue$.dequeue()
\STATE \hspace{3mm} {\bf foreach} $un$ in unvisited-neighbors($current$-$node$)
\STATE \hspace{6mm} {\bf while} (($current$-$node$ and $un$) not satisfy the
\STATE \hspace{6mm} $k$-anonymity non-reconstructability condition)
\STATE \hspace{9mm} change($un$)
\STATE \hspace{9mm} {\bf if} (any previously-satisfied $k$-anonymity
\STATE \hspace{9mm} non-reconstructability condition violates
\STATE \hspace{9mm}  between $un$ and visited- neighbors($un$))
\STATE \hspace{12mm} call change($un$) until there is no violation
\STATE \hspace{12mm} between $un$ and visisted-neighbors($un$)
\STATE \hspace{6mm} visit($current$-$node$)
\STATE \hspace{6mm} $processing$-$queue$.enqueue($un$)

\end{algorithmic}
\end{algorithm}

Algorithm \ref{alg:deltaselectivity}  shows how $\delta$-selectivity
can be enforced on a fragmentation ${\cal F}$. This algorithm is
very similar to Algorithm \ref{alg:dgbe}. It is worth noting that
the class values of equivalence classes in the dependency graph
({\it $d$-$graph$}) are at the  equivalence class-level. Subroutine $change$ simply removes a class value in an equivalence class. In order to reduce the class value distortion, the class value having the lowest frequency before converting the class values into equivalence class-level must be removed.

\begin{algorithm}
\caption{$\delta$-selectivity($d$-$graph$, $\delta$)}          
\label{alg:deltaselectivity}                           
\begin{algorithmic}[1]                    
\STATE $processing$-$queue$ = {\bf empty} \STATE mark all nodes in
$d$-$graph$ as $unvisited$ \STATE
$processing$-$queue$.enqueue(random($d$-$graph$)) \STATE {\bf while}
($processing$-$queue$ {\bf not empty}) \STATE \hspace{3mm}
$current$-$node$ = $processing$-$queue$.dequeue() \STATE
\hspace{3mm} {\bf foreach} $un$ in
unvisited-neighbors($current$-$node$) \STATE \hspace{6mm}  {\bf
while} ($\eta$($un$,$current$-$node$)$<$$\delta$) \STATE
\hspace{9mm} change($un$) \STATE \hspace{9mm} {\bf if} ($\eta$
between $un$ and any $vn\in$ visited- \STATE \hspace{9mm} neighbors($un$)
becomes less  than $\delta$)
\STATE \hspace{12mm} call change($un$) until $\eta$($un$,$vn$) becomes greater \STATE \hspace{12mm} than $\delta$
\STATE \hspace{6mm} visit($current$-$node$)
\STATE \hspace{6mm} $processing$-$queue$.enqueue($un$)

\end{algorithmic}
\end{algorithm}

\section*{Appendix B. Proofs}

\noindent THEOREM 4.1. {\em ($K$-ANONYMITY NON-RECONSTRUCTABILITY CONDITION). The condition for a given fragmentation  ${\cal F}=\{F_1,F_2\}$
which satisfies the fragment $k$-anonymity condition, and fragments
$F_1=\{ EQ_{11},$ $ EQ_{12}$ $ ,..., EQ_{1n} \}$ and  $F_2=\{
EQ_{21}, EQ_{22},$ $ ..., EQ_{2m} \}$, to be non-reconstructible is
that one of the following must be true for each joined  pair
$EQ_{1i}$, $EQ_{2j}$:
\squishlist
\item$\sum_{c \in C_{1i} \cap C_{2j}} freq(c,EQ_{1i})\times
freq(c,EQ_{2j}) =0$
\item $\sum_{c \in C_{1i} \cap C_{2j}} freq(c,EQ_{1i})\times
freq(c,EQ_{2j}) \geq k$
\squishend}

\begin{proof}
According to Definition \ref{def:nonreconstructable}, ${\cal F}$ is
$non$-$reconstructible$ if and only if the relation  $F_1 \Join F_2$
satisfies  $k$-anonymity.

Joining two equivalence classes $EQ_{1i}$ and $EQ_{2j}$ generates
tuples with the same value for their quasi-identifiers. If the
number of generated tuples is greater than $k$, joining $EQ_{1i}$
and $EQ_{2j}$ does not violate the $k$-anonymity property. For any
given equivalence classes $EQ_{1i}$ and $EQ_{2j}$, one of the
following conditions is true:

\squishlist

\item  $CLS = C_{1i} \cap C_{2j} =\emptyset$.
\item  $CLS = C_{1i} \cap C_{2j}= \{c_1$ $,c_2,...,c_w\}$.

\squishend

If  the former condition is true, $EQ_{1i}$ and $EQ_{2j}$ do not
have any common values to be joined on.  Therefore, we have:
\[ \sum_{c \in \emptyset} freq(c,EQ_{1i})\times
freq(c,EQ_{2j}) = 0
\]
 If the latter condition is true, the joining of  $EQ_{1i}$ and $EQ_{2j}$
 will result in $\sum_{c \in CLS} freq(c,EQ_{1i}) \times
freq(c,EQ_{2j})$ tuples. Thus, the result of joining $EQ_{1i}$ and
$EQ_{2j}$ is $k$-anonymous if the number of resulting tuples is at
least $k$.  Therefore, in either case the resulting fragmentation
${\cal F}$ is $k$-anonymity non-reconstructible.
\end{proof}

\noindent THEOREM 5.1. {\em  The data set resulting from
joining chunks $CK_{s1}$ and $CK_{s2}$ neither violates
$k$-anonymity nor $\ell$-diversity.}

\begin{proof} To prove this theorem,
we discriminate between different types of $\ell$-diversity, and
investigate each model individually.

\vspace{5mm}

{\bf Distinct $\ell$-diversity}:  The joining of  $CK_{s1}$ and
$CK_{s2}$ is performed based on joining equivalence classes
$EQ_{i1}^s$ and $EQ_{j2}^s$. According to the Step 4 of
fragmentation-based $\ell$-diversity, $EQ_{i1}^s$ and $EQ_{j2}^s$
have been constructed in such a way that have the same set (of size
$l_s$) of class values with frequencies
$\{f_1^x,f_2^x,...,f_{l_s}^x\}$ and $\{f_1^y,f_2^y,...,f_{l_s}^y\}$,
respectively. Since both $EQ_{i1}^s$ and $EQ_{j2}^s$ satisfy the
$k$-anonymity, we have:

\vspace{3mm}

$1) |EQ_{i1}^s| = f_1^1+f_2^1+...+f_{l_s}^1 \geq k$

$2) |EQ_{j2}^s| = f_1^2+f_2^2+...+f_{l_s}^2 \geq k$

\vspace{3mm}

Evidently, joining $EQ_{i1}^s$ and $EQ_{j2}^s$ are based on the
common sensitive values, thus it results in a data set of size
$f_1^1 \times f_1^2+f_2^1 \times f_2^2+...+f_{l_s}^1 \times
f_{l_s}^2$ having exactly $l_s$ sensitive values. Therefore, the
data set resulting from this join satisfies the distinct
$\ell$-diversity. Since joining each pair of arbitrary equivalence
classes $EQ_{i1}^s$ and $EQ_{j2}^s$ satisfies the distinct
$\ell$-diversity, joining  $CK_{s1}$ and $CK_{s2}$ also satisfies this
privacy requirement.

{\bf Entropy (or recursive) $\ell$-diversity}: This part is a bit
trickier. Joining each pair of equivalence classes $EQ_{i1}^s$ and
$EQ_{j2}^s$ will result
 in a data set with at least $k$ tuples (with the same reasoning as distinct $\ell$-diversity),
 but most likely different level of diversity than $l_s$ (it can be lower or higher than $l_s$).
 This might be considered as a privacy violation. However, joining all possible pairs of equivalence
 classes $EQ_{i1}^s$ and $EQ_{j2}^s$ generates a data set consisting of both
 real (all tuples in $S_s$) and fake tuples. There is an important observation here.
 If the attacker has enough background knowledge to rule out the faked tuples,
 the diversity level of real tuples are exactly equal to $l_s$. Hence, this is not a privacy violation.

\end{proof}

\section*{Appendix C. Enforcement via $\delta$-selectivity }
\section*{I. Tuple-level vs. equivalence class-level class value publishing}
Table ~\ref{tuplelevelvsequilevel} exemplifies the difference
between  tuple-level and  equivalence class-level  publishing for a
 5-anonymized fragment. As Table ~\ref{tuplelevelvsequilevel}b
illustrates, publishing the class values in the equivalence
class-level usually results in an equivalence class with some {\it
ambiguous slots} for its class values. For example,   two ambiguous
slots are illustrated in Table \ref{tuplelevelvsequilevel}b. The
only case in which publishing at the equivalence class-level does
not result in an ambiguous slot is when the class values are unique
in the equivalence class. As no further information is published
regarding the frequency of the class values in each equivalence
class, any class value available in the equivalence class can be
placed in the ambiguous slots.
\vspace{-2mm}
\begin{table}[h]
\centering
\caption{Tuple-level vs. equivalence class-level class value publishing }
\label{tuplelevelvsequilevel}
\subfloat[Tuple-level class value publishing for $k=5$.]{
\begin{tabular}{|l c|c|}
\hline
{\tiny 23} & {\tiny M}  & {\tiny flu} \\ \cline{3-3}
{\tiny 23} & {\tiny M}  & {\tiny pneumonia} \\ \cline{3-3}
{\tiny 23} & {\tiny M}  & {\tiny dyspepsia} \\ \cline{3-3}
{\tiny 23} & {\tiny M}  & {\tiny pneumonia} \\ \cline{3-3}
{\tiny 23} & {\tiny M}  & {\tiny flu} \\
\hline
\end{tabular}
}
\hspace{20mm}
\subfloat[Equivalence class-level class value publishing for $k=5$.]{
\begin{tabular}{|l c|c|}
\hline
{\tiny 23} & {\tiny M}  & {\tiny } \\
{\tiny 23} & {\tiny M}  & {\tiny flu} \\
{\tiny 23} & {\tiny M}  & {\tiny pneumonia} \\
{\tiny 23} & {\tiny M}  & {\tiny dyspepsia} \\
{\tiny 23} & {\tiny M}  & {\tiny } \\
\hline
\end{tabular}
}\end{table}

The  ambiguous slots in an equivalence class
$EQ_{ij}$, published at the  equivalence class-level, can take any
of the class values in $C_{ij}$. In other words, different versions
for $EQ_{ij}$ can be assumed. Table \ref{table:versions} illustrates
two possible versions for the equivalence class in Table
\ref{tuplelevelvsequilevel}b.

\begin{table}[h]
\centering
\caption{Two possible versions of Table \ref{tuplelevelvsequilevel}b }
\label{table:versions}
\subfloat[Version 1]{
\begin{tabular}{|l c|c|}
\hline
{\tiny 23} & {\tiny M}  & {\tiny flu} \\ \cline{3-3}
{\tiny 23} & {\tiny M}  & {\tiny flu} \\ \cline{3-3}
{\tiny 23} & {\tiny M}  & {\tiny dyspepsia} \\ \cline{3-3}
{\tiny 23} & {\tiny M}  & {\tiny pneumonia} \\ \cline{3-3}
{\tiny 23} & {\tiny M}  & {\tiny flu} \\
\hline
\end{tabular}
}
\hspace{20mm}
\subfloat[Version 2]{
\begin{tabular}{|l c|c|}
\hline
{\tiny 23} & {\tiny M}  & {\tiny pneumonia} \\ \cline{3-3}
{\tiny 23} & {\tiny M}  & {\tiny flu} \\ \cline{3-3}
{\tiny 23} & {\tiny M}  & {\tiny dyspepsia} \\ \cline{3-3}
{\tiny 23} & {\tiny M}  & {\tiny pneumonia} \\ \cline{3-3}
{\tiny 23} & {\tiny M}  & {\tiny dyspepsia} \\
\hline
\end{tabular}
}\end{table}

\section*{II. A subtle way to find $\eta$}
Given two equivalence classes $EQ_{1i}$ and $EQ_{2j}$ whose class
values are published in equivalence class-level and have at least
one class value in common, we find the number of tuples resulted
from their join which is referred to as {\it equijoin selectivity}
in the literature. As $|EQ_{1i}|$ and $|C_{1i}|$ refer to the number
of tuples and the class values (in equivalence class-level) in
$EQ_{1i}$, the number of ambiguous slots is equal to $|EQ_{1i}|
-|C_{1i}|$.
 The set $\{V(EQ_{1i})\}$ represents different versions (or instantiations)
 of $EQ_{1i}$. Each member  of this set is a combination
of $|EQ_{1i}|$ tuples in which there exist at least one tuple per
each class value in $C_{1i}$. The size of this set is equal to
${(|EQ_{1i}|-|C_{1i}|)+|C_{1i}|-1} \choose {(|EQ_{1i}|-|C_{1i}|)}$.

 There exist
$|\{V(EQ_{1i})\}| \times |\{V(EQ_{2j})\}|$  different ways to join
$EQ_{1i}$ and $EQ_{2j}$. Among all possible joins, those generating
$k$ tuples (or more)  are referred to as {\it
$k$-anonymity-preserving equijoins}.

In $\eta$ formula, calculating the denominator is
straightforward by multiplying  $|{V(EQ_{1i})}|$ and
$|{V(EQ_{2j})}|$. However, computing the numerator is more
challenging.  The simplest way is perform all joins, and count those
which satisfy the minimum cardinality of $k$. A more subtle way is
to consider the conditions (constraints) that exist on the number of
class values in each equivalence class and map the problem as
follows:

The number of pairs ($A$,$B$) of integer vectors
$A=$($a_1,a_2,...,a_c$) and $B=$($b_1,b_2,...,b_c$) in which $a_d$
and $b_d$ depict the frequency of $d^{th}$  common class value in
$V(EQ_{1i})$ and $V(EQ_{2j})$ respectively, such that the following
conditions are satisfied:


\hspace{2mm}

\noindent condition 1: $\sum{a_d\times b_d} \geq k $


\noindent condition 2: $1 \leq a_d \leq (1+|EQ_{1i}|-|C_{1i}|)$


\noindent condition 3: $1 \leq b_d \leq (1+|EQ_{2j}|-|C_{2j}|)$


\noindent condition 4: $\sum {a_d} \leq |CLS|+|EQ_{1i}|-|C_{1i}|$


\noindent condition 5: $\sum {b_d} \leq |CLS|+|EQ_{2j}|-|C_{2j}|$


\noindent where $CLS=C_{1i} \cap C_{2j}$.

\section*{Appendix D. Experiments}
\section*{I. Performance measures}
We utilized two metrics to evaluate the effectiveness of our
proposed method. The first one is the  widely-used metric known as
{\em information loss}, which   captures the total amount of lost
information due to generalization. In fact, this metric shows
the usefulness of data {\em on a per-attribute basis}  for general
workloads. For an anonymized data set with $n$ tuples and $m$
attributes, the information loss  $I$ is computed  as follows:

$I= \sum_{i=1}^{n} \sum_{j=1}^{m}\frac{|upper_{ij}-lower_{ij}|}{n
\cdot m \cdot |max_j-min_j|}$

Here,  $lower_{ij}$ and  $upper_{ij}$ represent  the lower and upper
bounds of attribute $j$ in tuple $i$ after generalization, and
$min_j$ and $max_j$ represent the maximum and minimum values taken
by attribute $j$  over all records. Note that the computation in the
fragmented and unfragmented case is not different, as long as all
the different attributes in the different fragments are used. This
is the most direct measure of the data quality, since it normalizes
the final result by the number of attributes.

To evaluate the utility of the data anonymized by our
meta-algorithm, we calculated the weighted F-measure of a classifier
trained on the anonymous fragmentation. This metric reflects the
goodness of classifier more accurately in case on unevenly
distributed test data. Each fragment was used to train the
classifier separately. For a given test instance,  the different
fragments of the training data were trained separately, and the
weighted majority label from the different classifiers was reported,
where the weight used was score returned by the classifier from each
fragment. Each class label was once considered as positive (and once
as negative) and the weighted F-measure for each case was calculated
by taking into account the fraction of the positive instances. We
used J48\footnote{J48 is an open source implementation of C4.5 in
Java,
\url{http://weka.sourceforge.net/doc/weka/classifiers/trees/J48.html.}}
and a
$k$-NN\footnote{\url{http://weka.sourceforge.net/doc/weka/classifiers/lazy/IBk.html.}}
classifier in Weka with the default setting. The value of $k$ in
$k$-NN was set to 5. The learning from generalized values was  also
done by the technique used in \cite{fevre2}. In each case, the
decomposition was performed into two fragments.

In addition, the amount of distortion required to the class values (for fragmentation $k$-anonymity)
was measured. Specifically, the number of distorted class values
were computed for each of the different techniques. The aim is to
show that the amount of distortion required was relatively small.

\section*{II. Data set description}
Real  data set \textit{Musk} 
from the {\em  UCI Machine Learning
Repository}\footnote{\url{http://archive.ics.uci.edu/ml.}} was
used.  It contains 7074 instances with 168
numerical feature attributes describing a set of molecules. The goal
is to predict whether a molecule is musk or non-musk.  In order to
test the effects  of varying data dimensionality, we  chose 10, 20,
30, and 40 features randomly and constructed four versions of $Musk$
with different dimensionality. Around $70\%$ of
the $Musk$ data set was used for training. The distribution  of the
two classes  in the training and test data set is shown in Table
\ref{table:datasetdis}.  Although the data set used in this work is numerical, the proposed meta-algorithm can be used to anonymize categorical values by simply using an off-the-shell anonymization technique capable to anonymize categorical values in Step 2 and 4 of fragmentation $k$-anonymity and $\ell$-diversity, respectively.

\begin{table}
\caption{Class distributions in $Musk$ data set} \label{table:datasetdis} \centering
\begin{tabular}{|c|c|c|}
\hline
{\scriptsize {\bf class}} & {\scriptsize {\bf train}} & {\scriptsize {\bf test}} \\
\hline

{\scriptsize musk} & {\scriptsize 1017} & {\scriptsize 207 } \\
\hline

{\scriptsize non-musk} & {\scriptsize 3983 } & {\scriptsize 1867} \\
\hline
\end{tabular}

\end{table}

\end{document}